\documentclass[10pt,journal,compsoc]{IEEEtran}

\ifCLASSOPTIONcompsoc
  \usepackage[nocompress]{cite}
\else
  \usepackage{cite}
\fi

\ifCLASSINFOpdf
   \usepackage[pdftex]{graphicx}
   \graphicspath{{figure/}}
   \DeclareGraphicsExtensions{.pdf,.jpeg,.png}
\else
   \usepackage[dvips]{graphicx}
   \graphicspath{{../eps/}}
   \DeclareGraphicsExtensions{.eps}
\fi

\usepackage{amsmath}
\usepackage{algorithmic}
\usepackage{threeparttable}
\usepackage{multirow}
\usepackage{amssymb}
\usepackage{pifont}
\usepackage{arydshln}
\usepackage{afterpage}

\usepackage{float}

\ifCLASSOPTIONcompsoc
  \usepackage[caption=false,font=footnotesize,labelfont=sf,textfont=sf]{subfig}
\else
  \usepackage[caption=false,font=footnotesize]{subfig}
\fi

\begin{document}

\title{DarwinWafer: A Wafer-Scale Neuromorphic Chip}

\author{%
  Xiaolei Zhu\textsuperscript{1,2,*},
  Xiaofei Jin\textsuperscript{2,3,*},
  Ziyang Kang\textsuperscript{3},
  Chonghui Sun\textsuperscript{1},
  Junjie Feng\textsuperscript{2,4},
  Dingwen Hu\textsuperscript{1,2},
  Zengyi Wang\textsuperscript{1},
  Hanyue Zhuang\textsuperscript{1},
  Qian Zheng\textsuperscript{2,3},
  Huajin Tang\textsuperscript{2,3},
  Shi Gu\textsuperscript{2,3},
  Xin Du\textsuperscript{2},
  De Ma\textsuperscript{2,3,\textdagger},
  and~Gang Pan\textsuperscript{2,3,\textdagger}%
  \IEEEcompsocitemizethanks{%
    \IEEEcompsocthanksitem * Equal contribution (co-first authors).
    \IEEEcompsocthanksitem \textdagger~Corresponding authors. E-mail: made@zju.edu.cn; gpan@zju.edu.cn.
    \IEEEcompsocthanksitem \textsuperscript{1} College of Integrated Circuits, Zhejiang University, Hangzhou 310027, China.
    \IEEEcompsocthanksitem \textsuperscript{2} State Key Laboratory of Brain-Machine Intelligence, Zhejiang University, Hangzhou 310027, China.
    \IEEEcompsocthanksitem \textsuperscript{3} College of Computer Science and Technology, Zhejiang University, Hangzhou 310027, China.
    \IEEEcompsocthanksitem \textsuperscript{4} College of Information Science \& Electronic Engineering, Zhejiang University, Hangzhou 310027, China.
    \IEEEcompsocthanksitem This work is supported by the National Key Research and Development Program of China (No. 2024YDLN0005), the Major Scientific Research Project of Zhejiang Province (No.2022C01048), the Pre-research project (No.31513010501), the NSFC of Zhejiang Province (LDT23F0401), the Sino-German Mobility Programme (No. M-0499) and Natural Science Foundation of China (No. 61925603).
  }%
}

\IEEEtitleabstractindextext{
\begin{abstract}
Neuromorphic computing promises brain-like efficiency, yet today’s multi-chip systems scale over PCBs and incur orders-of-magnitude penalties in bandwidth, latency, and energy, undermining biological algorithms and system efficiency. We present DarwinWafer, a hyperscale system-on-wafer that replaces off-chip interconnects with wafer-scale, high-density integration of 64 Darwin3 chiplets on a 300 mm silicon interposer. A GALS NoC within each chiplet and an AER-based asynchronous wafer fabric with hierarchical time-step synchronization provide low-latency, coherent operation across the wafer. Each chiplet implements 2.35 M neurons and 0.1 B synapses, yielding 0.15 B neurons and 6.4 B synapses per wafer. At 333 MHz and 0.8 V, DarwinWafer consumes ~100 W and achieves 4.9 pJ/SOP, with 64 TSOPS peak throughput (0.64 TSOPS/W). Realization is enabled by a holistic chiplet–interposer co-design flow (including an in-house interposer-bump planner with early SI/PI and electro-thermal closure) and a warpage-tolerant assembly that fans out I/O via PCBlets and compliant pogo-pin connections, enabling robust, demountable wafer-to-board integration. Measurements confirm $\leq$
10 mV supply droop and a uniform thermal profile (34–36 °C) under ~100 W. Application studies demonstrate whole-brain simulations: two zebrafish brains per chiplet with high connectivity fidelity (Spearman r = 0.896) and a mouse brain mapped across 32 chiplets (r = 0.645). To our knowledge, DarwinWafer represents a pioneering demonstration of wafer-scale neuromorphic computing, establishing a viable and scalable path toward large-scale, brain-like computation on silicon by replacing PCB-level interconnects with high-density, on-wafer integration.

\end{abstract}

}

\maketitle

\IEEEdisplaynontitleabstractindextext

\IEEEpeerreviewmaketitle

\IEEEraisesectionheading{\section{Introduction}\label{sec:introduction}}

\IEEEPARstart{T}{he} human brain, a biological organ consuming a mere 20\,watts \cite{lennie2003cost}, exhibits cognitive, learning, and creative capabilities that far surpass those of today's most powerful supercomputers. This profound efficiency gap reveals the fundamental limitations of the conventional von Neumann architecture \cite{backus1978programming} and has inspired a decades-long quest to build brain-like neuromorphic hardware \cite{mead1990neuromorphic}. This endeavor aims not only to provide neuroscientists with unprecedented tools to explore the mysteries of the brain but also to offer a more energy-efficient and sustainable path toward solving real-world intelligence problems.

In pursuit of this vision, academia and industry have developed numerous landmark neuromorphic hardware platforms. These include industry-led systems like Intel's Loihi \cite{davies2018loihi,intel_loihi2} and IBM's TrueNorth \cite{merolla2014million}, as well as academic platforms such as SpiNNaker \cite{furber2014spinnaker,mayr2019spinnaker2}, BrainScaleS \cite{schmitt2017brainscales,pehle2022brainscales2,schemmel2010wafer}, Tianjic \cite{pei2019tianjic} and the Darwin series of neuromorphic processors\cite{ma2024darwin3,Ma2017Darwin}. These systems have validated the immense potential of neuromorphic computing at the chip level. A common design philosophy underpinning these systems is scalability, typically achieved by interconnecting multiple chips on a printed-circuit board (PCB) \cite{arunkumar2017mcmgpu,shao2019simba} to construct larger neural networks. This approach, however, represents the fundamental bottleneck and core paradox of the current scaling paradigm.

The intelligence of the human brain emerges from its immense scale (\(\sim10^{11}\) neurons) and dense connectivity (\(\sim10^{15}\) synapses) \cite{herculano2009numbers}, a scale effect that presents an insurmountable chasm for current PCB-based scaling strategies. Interconnecting individually packaged neuromorphic chips via a PCB degrades inter-chip communication bandwidth, latency, and energy efficiency by several orders of magnitude compared to on-chip communication \cite{shulaker2017integration}. This is analogous to fitting a set of high-speed processors with exceedingly slow “data pipes”. As the system scales, inefficient off-chip communication not only offsets most of the energy-efficiency gains achieved on-chip but also alters the dynamics of Spike-Timing-Dependent Plasticity (STDP) algorithms, thereby compromising their biological plausibility \cite{pfeiffer2018spiking}. Consequently, while individual chips are highly advanced, constructing a system comparable in scale to the human brain would result in a "behemoth" composed of tens of thousands of chips, occupying an enormous volume and consuming staggering amounts of power. This hardware scaling challenge is intensified by the growing complexity and scale of modern Spiking Neural Network (SNN) models, such as deep residual networks \cite{Hu2023SpikingResNet} and those converted from conventional ANNs \cite{Hu2023FastSNN}, which demand unprecedented computational resources \cite{Hu2024LargeScaleSNN}. The current scaling path is, at a physical level, antithetical to the brain’s design philosophy of dense and efficient integration.

\begin{figure*}[!h]
    \centering
    \includegraphics[width=\linewidth]{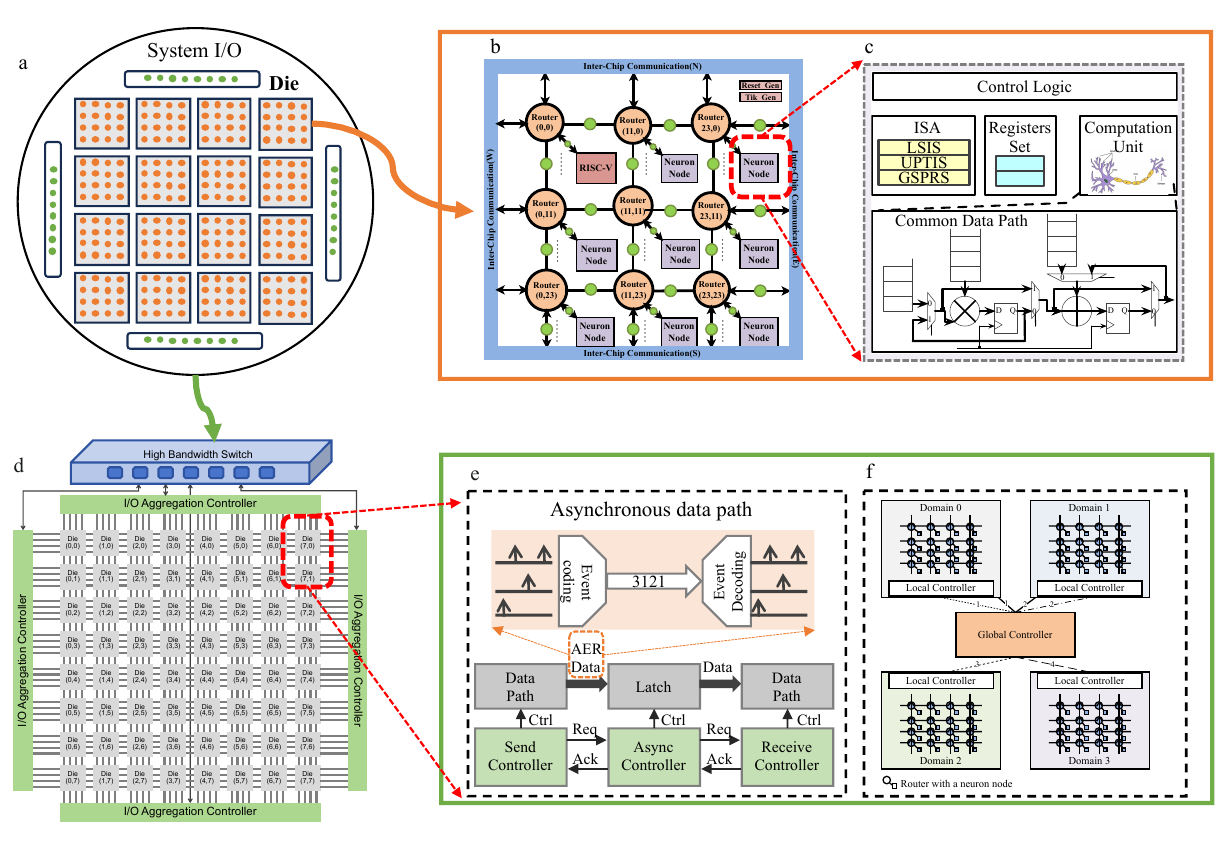}
    \caption{ Hierarchical architecture and operational mechanisms of the DarwinWafer System-on-Wafer (SoW). 
    (a) Overview of the DarwinWafer architecture. The figure shows wafer-scale integration of a large array of neuromorphic dies on a single substrate, with dedicated system I/O at the periphery. 
    (b) The internal architecture of a single Darwin3 die, which serves as the fundamental building block. It features a mesh-based NoC interconnecting a RISC-V control core with numerous neuromorphic neuron nodes. 
    (c) The microarchitecture of each Neuron Node, which is built upon a custom Instruction Set Architecture (ISA) and features a specialized registers set for state variables and a computation unit for efficient neurodynamic processing. 
    (d) The logical organization of the wafer-scale, AER-based (Address-Event Representation) silicon interconnect architecture, where I/O Aggregation Controllers manage communication between the die array and an external high-bandwidth switch. 
    (e) The detailed mechanism for standardized AER communication, implemented with an asynchronous data path that uses event coding/decoding and a request/acknowledge handshake protocol for reliable, low-latency data transfer. 
    (f) The high-precision time-step synchronization mechanism, which employs a hierarchical structure featuring a global controller that coordinates multiple local controllers to ensure coherent, time-stepped operation across the entire parallel system.
    }
	\label{fig1}
\end{figure*}

To break this scaling–communication bottleneck, we introduce and demonstrate DarwinWafer, a hyperscale wafer-scale neuromorphic system. An overview of the wafer-scale architecture is shown in Fig.~\ref{fig1}(a). We posit that superior scalability can be achieved through the high-density integration of chiplets at the physical scale of a wafer, compared to relying on inefficient off-chip interconnects. This approach aligns with a recent resurgence of interest in wafer-scale computing, validated by commercial systems from industry leaders \cite{talpes2022dojo,lie2022cerebras}, and is built upon the foundation of mature chiplet-based design paradigms and advanced 2.5D packaging technologies. By leveraging these techniques in conjunction with our holistic co-design methodology, DarwinWafer integrates 64 Darwin3 \cite{ma2024darwin3} neuromorphic chiplets via flip-chip bonding onto a single 300 mm wafer-scale silicon interposer with high-density interconnects. Each Darwin3 chiplet contains 2.35 million neurons and 0.1 billion synapses, bringing the 

\noindent total integrated neurons and synapses count on the wafer-scale chip DarwinWafer to 0.15 billion and 6.4 billion respectively. Measurement results show that operating at a frequency of 333 MHz with a 0.8 V supply voltage, the entire DarwinWafer has a typical power consumption of around 100 W with an energy efficiency of 4.9 pJ per synaptic operation (SOP). This wafer-scale integration method effectively alleviates the communication bottleneck between neuromorphic chiplets, achieving unprecedented integration density, communication bandwidth, and system-level energy efficiency, thereby paving a viable technological path toward the realization of a true “Brain-on-Silicon”.

\section{Wafer-Scale Interconnect Architecture}
\subsection{Darwin3 Chip Architecture} 
The architecture of the Darwin3 chiplet is meticulously engineered to serve as a highly efficient and scalable building block for wafer-scale neuromorphic systems. Its core design philosophy centers on a globally asynchronous, locally synchronous (GALS) paradigm, realized through a two-dimensional mesh-based network-on-chip (NoC) that interconnects a multitude of specialized neuromorphic computing cores.The internal organization of a single Darwin3 die is shown in Fig.~\ref{fig1}(b). This foundational architecture is supported by a series of key innovations in its computational core, memory organization, and network protocol, all designed to achieve a harmonious balance between biological fidelity, computational efficiency, and system scalability. At the heart of this architecture is a neuromorphic computing core built upon a custom Instruction Set Architecture (ISA), the culmination of extensive research into diverse neuron and synapse models, which enables both the flexible description of neurodynamics—including support for bio-plausible, reconfigurable neuron dynamics \cite{Xiao2025BioPlausibleNeuron}—and highly parallel operations for real-time state updates. The microarchitecture of each Neuron Node—including the ISA, state registers, and compute unit—is shown in Fig.~\ref{fig1}(c). To support the massive and complex connectivity inherent in brain-inspired networks, Darwin3 incorporates a novel synaptic memory architecture that maximizes storage utilization through topology-aware compression and a crucial mechanism for dynamic memory resource reallocation, adapting to the diverse demands of various applications. Ultimately, communication between all compute and memory units is realized through a 

\begin{figure*}[ht]
    \centering
    \includegraphics[width=\linewidth]{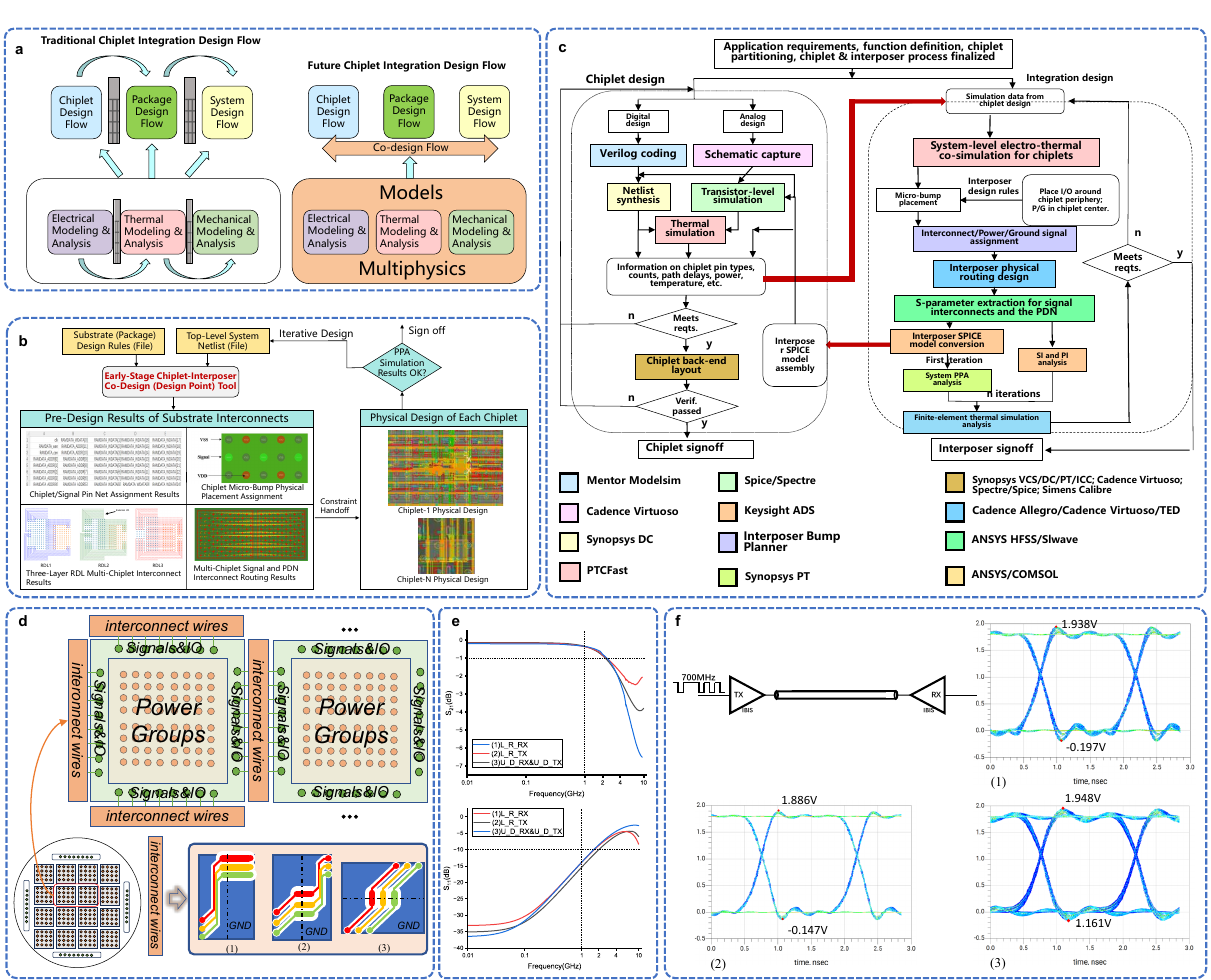}
    \caption{ Co-Design flow for DarwinWafer and simulation results. 
    (a) Traditional chiplet Integration design flow and Co-Design flow (adapted from \cite{HIR_ESTC2024}).
    (b) The proposed in-house IBPlanner (Interposer Bump Planner) tools.
    (c) The overall design flow and the EDA used in designing the DarwinWafer. 
    (d) “Signal-peripheral, power-central” design schematic, and three interconnect types:
    (1) L\_R\_RX: left-right interconnects for RX; 
    (2) L\_R\_TX: left-right interconnects for TX; 
    (3) U\_D\_RX and U\_D\_TX: up-down interconnects for RX/TX between dies.
    (e) S-parameter simulation validation for three types of interconnect wires in (d).
    (f) Eye diagrams at 700 MHz corresponding to structures (1), (2), and (3) in (d) respectively.}
	\label{fig2}
\end{figure*}

\noindent high-efficiency GALS interconnect architecture. This fabric leverages the intrinsic advantages of asynchronous circuits and is enhanced by a scalable NoC technology featuring relative addressing and internal hardware relaying, which provides flexible, alternative paths for long-range data transmission to effectively alleviate network congestion and significantly improve overall communication performance. The standardized AER datapath—with event coding/decoding and the request/acknowledge handshake for reliable, low-latency transfer—is summarized in Fig.~\ref{fig1}(e).

\subsection{System-on-Wafer (SoW) Design}
\subsubsection{Hierarchical Architecture for Scalable Integration}
The DarwinWafer SoW is architected upon a hierarchical framework designed for massive scalability, spanning from individual chiplets to the wafer and extending to multi-wafer clusters. At the foundational level, the system employs a silicon substrate interconnect technology (2.5D with TSVs) to directly bond rigorously screened neuromorphic chiplets, forming a high-density computational array. Communication within this array is facilitated by a scalable NoC, where reconfigurable internal synaptic routing information enables spike packets to traverse the entire wafer without requiring header modifications, thereby significantly reducing data-forwarding overhead. At the system interface, a dedicated off-wafer controller centralizes the management of peripheral signals, performing both bandwidth aggregation and crucial protocol conversion to standard interfaces like 10G/100G Ethernet.The wafer-scale AER fabric and the I/O aggregation path are illustrated in Fig.~\ref{fig1}(d). This allows a single SoW to function as a standard compute module within larger cluster systems. Critically, the asynchronous inter-chip interfaces decouple the clock domains of the individual chiplets. This design choice is a cornerstone of the system's flexibility, permitting the heterogeneous integration of chiplets from different process nodes, with varying yield levels, and operating at different frequencies, which vastly enhances the system's overall scalability and adaptability.

\subsubsection{Mechanisms for System-Level Reliability and Synchronization}
Ensuring the robust operation of such a highly integrated system requires a sophisticated approach to reliability and synchronization, as traditional methods of component replacement and external monitoring are unfeasible. To address this, we have developed a comprehensive reliability framework centered on three core strategies: minimizing on-wafer auxiliary components to reduce failure points, implementing an autonomous self-diagnosis and recovery mechanism, and establishing a hardware-software co-designed fault remediation system managed by the off-wafer controller. This is realized through a hierarchical monitoring architecture that operates at the die, wafer, and system levels, complemented by redundant nodes and chiplets for rapid fault recovery. Concurrently, to manage the computational synchrony across this vast parallel system, we introduce a hierarchical time-step synchronization mechanism. Following a "divide and conquer" strategy, the wafer is partitioned into multiple domains, each managed by a local time-step controller, with one configurable as a global master (see Fig.~\ref{fig1}(f)). This structure is enhanced by an adaptive time-step method inspired by DarwinSync \cite{jin2025darwinsync} , which dynamically adjusts the time-step length through a master-slave coordination mechanism. This ensures that the entire system can maintain precise, nanosecond-level synchronization tailored to the dynamic complexity of diverse computational tasks, guaranteeing both computational accuracy and operational efficiency.

\section{Co-design flow}

The immense complexity of wafer-scale systems like DarwinWafer renders conventional, linear design flows inadequate, mandating a holistic co-design approach, as illustrated in the Fig. 2a. These siloed methodologies precipitate a triad of fundamental challenges: they encourage local optimizations that manifest as global failures, such as thermal crosstalk and signal integrity collapse; they exacerbate late-stage conflicts between a chiplet’s microscopic design constraints and the substrate’s macroscopic routability; and finally, they prolong design cycles by delaying the discovery of critical physical issues until costly, late-stage simulations or physical prototyping, making the entire process high-risk and difficult to converge\cite{zhao2025toward3dic,kim2019codesign25d}.

To address these challenges, we propose a holistic co-design methodology centered around an integrated Electronic Design Automation (EDA) flow, as illustrated in Fig. 2c. This framework partitions the entire DarwinWafer design process into three phases: System Design, Chiplet Design, and Integration Design. Each phase leverages a combination of open-source, in-house-developed, and commercial EDA tools to achieve a globally optimized design. The design flow mainly comprises the following three stages:

(1)	System Design Phase:

This initial phase aims to establish a physically aware system blueprint. We perform top-level partitioning of the wafer-scale neuromorphic system, defining the function, communication protocols, and key performance metrics for each Darwin3 chiplet, while concurrently selecting the optimal semiconductor process technologies for both the chiplets and the substrate.

(2)	Chiplet Design Phase:

This phase is dedicated to ensuring robust chiplet performance in a realistic system environment by breaking the assumptions of idealized design. It begins with preliminary design and interface definition, where each chiplet team develops initial circuit designs (e.g., in Verilog or Cadence Virtuoso). The key deliverable is a feedforward of a “physical interface specification” to the integration designers, which includes estimated power, pin count and types, and critical path delays. This specification provides the essential initial constraints for the physical interconnect design on the substrate.

The process then enters the core of our co-design methodology: load-aware iterative performance conver- gence. The integration team, using the preliminary inter- face specifications, performs an initial substrate layout and routing, from which physical interconnect parasitics are extracted. These critical parameters are then feedback to the chiplet designers, initiating an iterative optimization loop driven by real-world physical data. Within this loop, chiplet designers import the parasitics into their simulation environments (e.g., Synopsys VCS/Primetime) to perform load- aware simulations and refine their circuits. The updated performance data is, in turn, fed back to the integration team for synchronous adjustments to the substrate design. This feed-forward, feedback, and optimization cycle is repeated until the chiplet’s front-end design converges under realistic physical constraints. This iterative process directly resolves the challenge of decoupled design spaces by forcing a unification of the microscopic circuit and macroscopic interposer /packaging domains.

\begin{figure*}[ht]
  \centering
  \includegraphics[width=\linewidth]{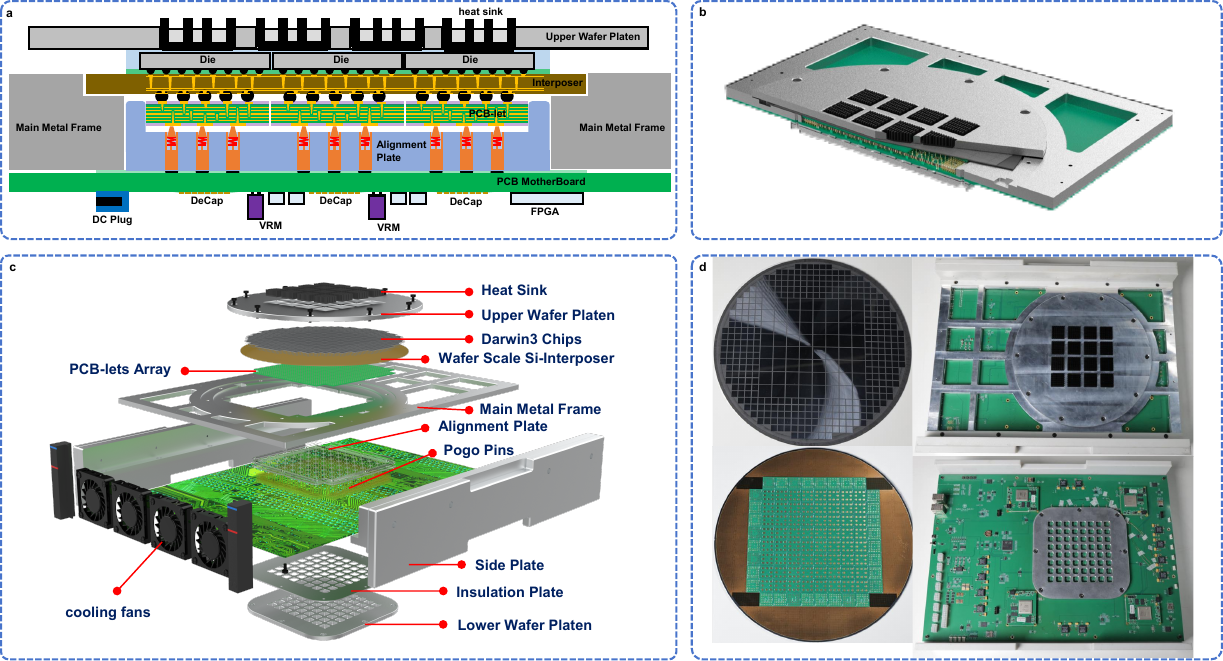}
  \caption[Fabrication flow of DarwinWafer]{
    Overall fabrication and assembling procedure of DarwinWafer. 
    (a) Assembled structure; 
    (b) cross-sectional view; 
    (c) exploded view;
    (d) Interposer, front and back: front populated with flip-chipped Darwin3 chiplets; back carrying soldered PCBlets.
  }
  \label{fig:darwin-flow}
\end{figure*}

(3)	Integration Design Phase:

The objective of this phase is the holistic co-verification and signoff of the system’s physical reliability, transforming the design blueprint into a robust physical entity. The phase commences with system-level electro-thermal simulation for the early design phase. Then using the converged interface data from the chiplets, we employ our in-house IBPlanner (Interposer Bump Planner) tools to complete the pin and microbump position allocation for the Darwin3 chiplet based on the front-end netlist design data output from the chiplet design phase. IBPlanner is a critical tool for enabling co-simulation of chiplets and interposers in the early design phase. As shown in Fig. 2c, IBPlanner uses the system-level top netlist and applies suitable algorithms, such as the Hungarian or Greedy algorithms, to determine possible interconnection paths between chiplets, sorting them by length. Based on the calculation of slack values between pins in the network, it assigns networks with specific slack values to the paths with the shortest lengths, thereby completing the allocation of pin and microbump positions for each chiplet. Combined with silicon interposer design rules, it determines the physical dimensions and shapes of the chiplets, completing the physical routing of the interposer. Using the physical dimension information of the interconnection lines between chiplets generated from the interposer pre-design, load information between chiplet pins can be obtained by extracting parasitic parameters. Feeding this information back into the physical design of each chiplet allows designers to obtain pin positions, dimensions, and load information at an early stage, effectively improving the efficiency and accuracy of the physical design.

Basically, the layout and routing for all 64 Darwin3 chiplets is guided by strategies such as “signal-peripheral, power-central,” (as shown in Fig. 2d) to create a floor plan and RDL design that systematically accounts for electrical, and timing constraints from the outset.

Following this initial integration, the system undergoes an iterative loop of cross-domain co-simulation, analysis, and refinement. We extract precise interconnect parameters to conduct system-level Signal Integrity (SI) and Power Integrity (PI) analyses (e.g., using Keysight ADS), concurrently with electro-thermal co-simulations using PTC- Fast. This comprehensive evaluation provides a ”full-system physical checkup.” Specifically, we designed three types of die-to-die interconnect wires on the substrate (as shown in Fig. 2d) to accommodate different interconnection requirements. The S-parameters of all three interconnect types fully met expectations: S11 remained below -10 dB at 1 GHz, while S21 exceeded -1 dB below 1 GHz as shown in Fig. 2e. Combined with the eye diagram results at 700 MHz (far exceeding practical requirements, shown in Fig. 2f), our interconnect design ensures nearly negligible signal reflection and attenuation during actual operation. Notably, the S-parameters of signal interconnects are converted to Spice models and fed back to the chiplet designers for final circuit signoff. If any metric fails to meet the design target, the system undergoes a targeted optimization cycle. By exposing and resolving physical-layer risks at this early stage through automated co-simulation, this process directly addresses the challenge of long and costly iteration cycles. After several iterations confirm that the system’s PPA meets the design objectives, we perform a final, high-fidelity thermal analysis using commercial-grade, finite-element-method-based tools (e.g., Icepak, COMSOL) for the ultimate signoff of the thermal management solution.

The integrated co-design process we proposed for the wafer-scale neuromorphic hardware system DarwinWafer couples chiplet design with system integration design, enabling collaboration between chiplet designers and integration designers at an early stage, significantly enhancing the overall performance of DarwinWafer and reducing design costs.

\section{Overall fabrication and assembling procedure} 
Following the electrical and thermal co-design of the Darwin3 chiplets and the wafer-scale system, the next critical step is to translate these design blueprints into physical reality. Although the Darwin3 chiplets can be fabricated with high yield using a standard 22nm SOI CMOS process, the fabrication and assembly of the wafer-scale system itself present a unique and formidable set of challenges that demand innovative solutions.

The core challenges we face stem from two primary sources. First is the difficulty of manufacturing a large-area, high-density wafer-scale interposer. This interposer must support die-to-die interconnects that span beyond standard reticle fields and integrate high-aspect-ratio Through-Silicon Vias (TSVs) for backside power and signal delivery. This requires advanced manufacturing techniques such as exposure field stitching and specialized TSV etching processes to ensure high-yield, reliable connections across the entire 300mm wafer. Second, and most critically, is the management of thermo-mechanical warpage during multi-level system assembly. The 2.5D flip-chip integration of 64 Darwin3 chiplets onto the silicon interposer creates a "die-to-wafer" composite. Due to the coefficient of thermal expansion (CTE) mismatch between silicon and the packaging materials, this composite exhibits significant warpage after high-temperature reflow processes. This warpage makes it nearly impossible to directly and reliably bond the large-area wafer composite to a conventional organic system mainboard (PCB), creating a major obstacle to achieving a compact and robust system-level interconnect.

To overcome these challenges, we propose an innovative hierarchical interconnection and assembly strategy for DarwinWafer. Our core insight is to decompose the problem by using a multi-level, transitional interconnect architecture and to absorb mechanical stress with an elastic, pressure-based connection mechanism. This strategy comprises two key innovations: first, the introduction of a PCBlet array for fan-out and density transition, and second, the use of pogo pins to achieve a compliant, elastic interconnection.

The physical implementation of this strategy is realized through a meticulously designed assembly procedure, as illustrated in Fig. 3. The process begins with the die-to-wafer composite, on which 64 Darwin3 chiplets will be flip-chipped onto the top side of the silicon interposer and the physical image shown in the upper part of Fig. 3d depicts 64 chiplets integrated onto the silicon interposer, with the small dies at the edge being dummy silicon chips attached to resist wafer warpage. On the backside of the interposer, an array of C4 bumps, connected via TSVs, is prepared for the next integration step. The first level of transitional interconnection involves bonding an array of PCBlets to the C4 bumps via a high-temperature reflow process. Within each PCBlet, its internal redistribution layers (RDLs) merge signals or power from multiple C4 bumps of the same net and fan them out to a grid of larger, wider-pitched contact pads on the PCBlet's backside. After processes including molding and thinning, a robust, sandwich-like die-to-wafer-to-PCBlet composite is formed as shown in the lower part of Fig. 3d which corresponds to the portion enclosed by the red solid-line polygonal frame in Fig. 3a.

\begin{figure*}[ht]
    \centering
    \includegraphics[width=\linewidth]{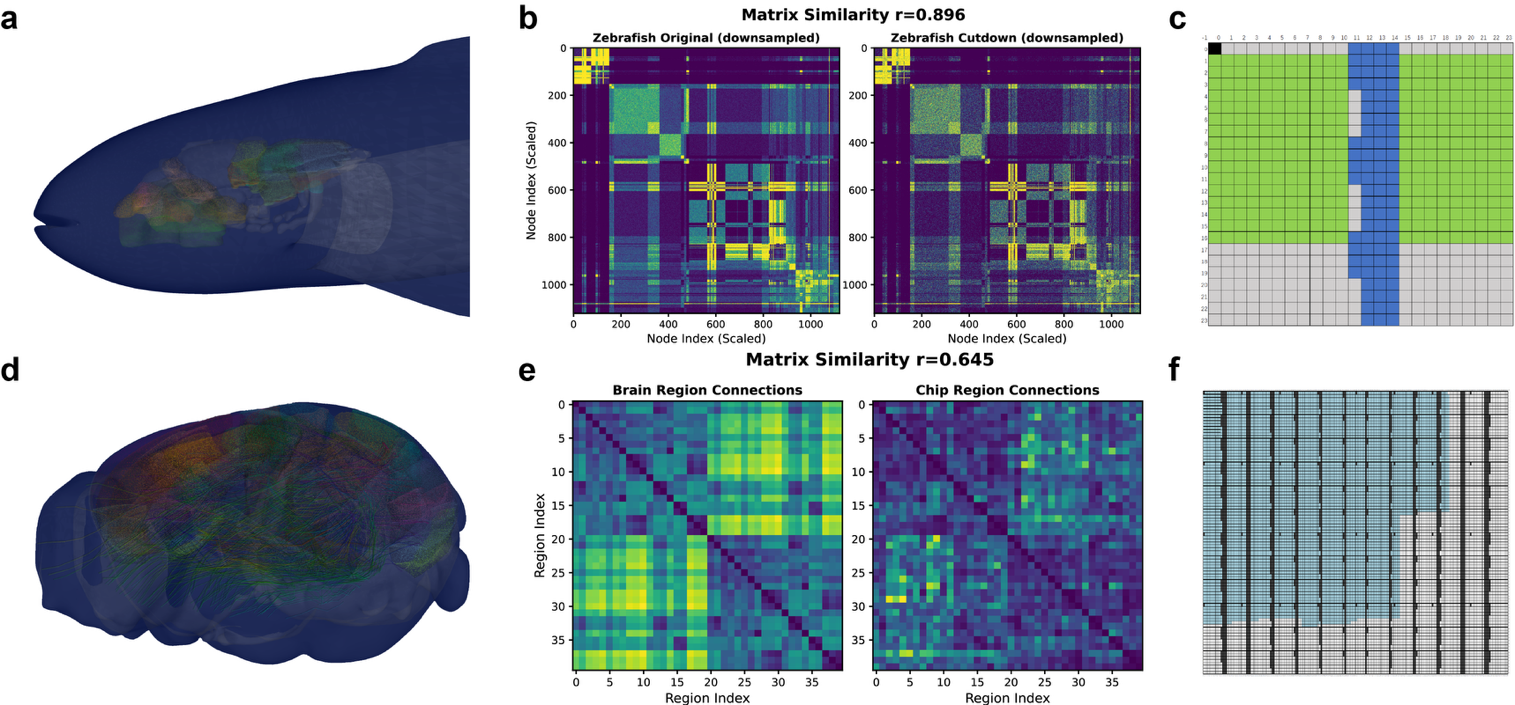}
    \caption{Simulation of zebrafish and mouse brains on DarwinWafer. 
    (a) Two zebrafish brains, each comprising ~70,000 neurons and ~640,000 synapses, are simulated on a single Darwin3 chip. 
    (b) The simulated connectivity matrix closely matches the biological one (Spearman correlation r = 0.896). 
    (c) Resource utilization: simulation components are colored green, system components blue. 
    (d) A single mouse brain (~9.5 million neurons, 500 million synapses) is simulated across multiple chips. 
    (e) The simulated connectivity matrix remains highly similar to the original (r = 0.645). 
    (f) The mouse brain simulation spans 32 chips; simulation utilisation is shown in cyan blue and system utilisation in black.}
    \label{fig:darwin_brain_sim} 
\end{figure*}

The second level of transitional interconnection connects this composite to the system mainboard. An array of pogo pins as well as a precision alignment plate are pre-soldered onto the mainboard as shown in the portion enclosed by the red dotted line polygonal frame in Fig. 3a. The die-to-wafer-to-PCBlet composite is then precisely pressed against the pogo pins array using a custom wafer clamp. The precision alignment plate with guide holes ensures that each pogo pins accurately mates with its corresponding PCBlet pad. The uniform pressure applied by the clamp causes the pogo pins to elastically deform, fully compensating for the height variations caused by warpage and establishing a reliable electrical connection across the entire wafer. For thermal management, the wafer clamp is designed with openings directly above each Darwin3 chiplet. Individual metal heat sinks can be mounted in these openings, making contact with the chiplets via a high-performance thermal interface material to ensure efficient heat dissipation for all 64 chiplets. After assembly as illustrated in Fig. 3a, we obtained the final structure of the overall SoW server system. Its cross-sectional view is presented in Fig. 3b, where the structure exactly corresponds to that of Fig. 3a. Expanding it into an exploded view configuration yields Fig. 3c.

This interconnection and assembly strategy for DarwinWafer is not only an effective solution to the warpage problem but also provides a generalizable framework for wafer-scale system integration. Its key advantages include insensitivity to warpage, demountability of the wafer composite from the mainboard (which greatly facilitates testing, debugging, and maintenance), and broad compatibility with various homogeneous or heterogeneous chiplet integration schemes.

\section{Measurement performance and application} 

\subsection{Measurement Performance of DarwinWafer}
\subsubsection{Physical and Electrical Characterization}
To characterize the fundamental physical and electrical properties of the DarwinWafer system, we developed a comprehensive testbed centered around a custom-built main board equipped with the pogo pins interconnect array. The system was powered by a programmable DC supply, allowing for precise voltage control and current monitoring. Power integrity was assessed by measuring the voltage droop on the power delivery network (PDN) under dynamically generated, high-intensity spike workloads. The thermal performance of the assembly and the per die heat sink solution was characterized by embedding thermal diodes within test chiplets and using a high-resolution infrared (IR) camera to capture the temperature distribution across the entire wafer surface during sustained operation. The physical characterization confirmed the robustness of our fabrication and assembly methodology. The power delivery network maintained a stable supply voltage with a maximum droop of only 10 mV under peak load conditions, indicating excellent power integrity. Thermally, the system exhibited a well-managed temperature profile, with a maximum measured die temperature of 36 °C and an average temperature of 34 °C across the wafer when dissipating a total power of 99.7 W. This demonstrates the effectiveness of our thermal management solution in handling the heat generated by 64 densely packed neuromorphic chiplets.

\subsubsection{Measured Performance}
The core computational and communication performance of DarwinWafer was evaluated using a suite of synthetic benchmarks and micro-kernels designed to stress specific architectural features. The system’s peak computational throughput was measured by executing highly parallelized neuron update and synaptic processing tasks across all 64 chiplets simultaneously. The total aggregated bisection bandwidth was benchmarked by creating traffic patterns that saturate the central communication links. System-level energy efficiency was determined by measuring the total power consumption from the main supply while the system was performing a known number of synaptic operations, allowing us to calculate the average energy per synaptic operation (SOP). The DarwinWafer system demonstrated state-of-the-art performance and efficiency. It achieved a peak computational throughput of 64 TSOPS, corresponding to 64 trillion synaptic operations per second. for worst-case paths. Most critically, the system achieved an exceptionally high energy efficiency, consuming an average of only 4.9 pJ per synaptic operation. This results in a system-level performance per watt metric of 0.64 TSOPS/W, validating our architecture’s capability to overcome the communication and energy efficiency bottlenecks inherent in traditional multi-chip systems.

\subsection{Brain simulations accelerated on DarwinWafer}
Brain simulations have emerged as a powerful tool for understanding the intricate structures and functions of neural networks. Here, to demonstrate the acceleration capacity of DarwinWafer to support efficient brain simulation, we presented simulations of zebrafish and mouse brains accelerated on the DarwinWafer platform with different system configurations (Fig. 4). In the case of the zebrafish brain, with its 70,000 neurons and 640,000 synapses, the simulation is conducted on DarwinWafer (Fig. 4a). The simulated connectivity matrix closely matches the original biological connectivity, with a high Spearman correlation of 0.896 (Fig. 4b), suggesting that the platform is effective at capturing the neural architecture of simpler organisms. This high correlation indicates that the DarwinWafer successfully reproduces the intricate wiring patterns of the zebrafish brain, highlighting its capacity to model smaller neural networks efficiently within the computational limits of a single chip. On the other hand, simulating the more complex mouse brain—comprising approximately 9.5 million neurons and 500 million synapses—requires the computational power of 32 chips (Fig. 4d). The connectivity matrix of the simulated mouse brain shows a Spearman correlation of 0.645 (Fig. 4e), reflecting a moderate degree of similarity to the original connectivity. While the correlation is lower than that of the zebrafish brain, it still demonstrates the potential of DarwinWafer to replicate the large-scale, more complex neural networks of mammals. This difference in correlation may arise due to the increased complexity of mammalian brain structures, which presents additional challenges in accurately simulating every aspect of their connectivity. Nonetheless, the simulation demonstrates that DarwinWafer can handle large, multi-chip configurations and replicate the broad structural patterns of more complex brains, making it a powerful hardware acceleration platform investigating higher-order neural processes. These simulations highlight the versatility and computational power of the DarwinWafer platform, as it adapts to simulate both small and large-scale brain models, efficiently allocating resources while maintaining high fidelity in the representation of biological connectivity.

\section{Conclusion} 
This work presented DarwinWafer, a hyperscale system-on-wafer neuromorphic platform that replaces PCB-scale interconnects with wafer-scale, high-density integration of 64 Darwin3 chiplets on a 300 mm silicon interposer. Through a GALS NoC per chiplet, an AER-based asynchronous wafer fabric, and hierarchical time-step synchronization, the system delivers coherent, low-latency operation across 0.15 B neurons and 6.4 B synapses. A holistic chiplet–interposer co-design flow—featuring early SI/PI and electro-thermal closure and an in-house interposer-bump planner—together with a warpage-tolerant PCBlet/pogo-pin assembly, enabled robust realization with measured $\leq$10 mV supply droop and a uniform 34–36 °C thermal profile at ~100 W. At 333 MHz, 0.8 V, DarwinWafer achieves 64 TSOPS peak throughput and 4.9 pJ/SOP (0.64 TSOPS/W), dissolving the scaling–communication bottleneck that constrains multi-chip neuromorphic systems. Application studies further demonstrate high-fidelity whole-brain simulations—from two zebrafish brains per chiplet (r = 0.896) to a mouse brain mapped across 32 chiplets (r = 0.645). These results establish wafer-scale neuromorphic integration as a viable path toward brain-like computation on silicon. Future work will extend to multi-wafer clusters, richer online plasticity and learning, tighter algorithm-architecture co-design and tooling, and heterogeneous technology integration to further advance capacity, efficiency, and robustness. Realizing the full potential of such wafer-scale hardware also necessitates a mature software ecosystem, including a comprehensive system software architecture \cite{Deng2022DarwinS} and efficient methodologies for mapping large-scale networks onto the physical hardware \cite{Jin2023MappingVLSNN}.

\ifCLASSOPTIONcaptionsoff
  \newpage
\fi

\bibliographystyle{IEEEtran}
\bibliography{IEEEabrv,bare_jrnl_compsoc}

\end{document}